\documentclass[12pt]{iopart}

\expandafter\let\csname equation*\endcsname\relax
\expandafter\let\csname endequation*\endcsname\relax
\usepackage{amsmath}
\usepackage{amsfonts}
\usepackage{amssymb}
\usepackage{graphicx}
\usepackage{subfigure}

\begin{document}

\title{Intrinsically narrowband pair photon generation in microstructured fibres}

\author{Alex Clark$^1$, Bryn Bell$^1$, J\'er\'emie Fulconis$^1$, Matth\"{a}us M. Halder$^1$, Ben Cemlyn$^1$, Olivier Alibart$^2$, Chunle Xiong$^3$, William J. Wadsworth$^3$, and John G. Rarity$^1$}
\address{$^1$ Centre for Communications Research, Department of Electrical and Electronic Engineering, University of Bristol, Merchant Venturers Building, Woodland Road, Bristol, BS8 1UB, UK}
\address{$^2$ Laboratoire de Physique de la Matière Condensée, Unité Mixte de Recherche 6622, Centre National de la Recherche Scientifique, Université de Nice-Sophia Antipolis, Parc Valrose 06108, Nice 2, France}
\address{$^3$ Centre for Photonics and Photonic Materials, Department of Physics, University of Bath, Claverton Down, Bath, BA2 7AY, UK}

\begin{abstract}
In this paper we study the tailoring of photon spectral properties generated by four-wave mixing in a birefringent photonic crystal fibre (PCF). The aim is to produce intrinsically narrow-band photons and hence to achieve high non-classical interference visibility and generate high fidelity entanglement without any requirement for spectral filtering, leading to high effective detection efficiencies. We show unfiltered Hong-Ou-Mandel interference visibilities of $77\%$ between photons from the same PCF, and $80\%$ between separate sources.  We compare results from modelling the PCF to these experiments and analyse photon purities.
\end{abstract}
\maketitle

\section{Introduction}

Single photons are at the heart of optical quantum technologies such as quantum cryptography \cite{gi-natphot-1-165}, quantum computing \cite{kn-nat-409-46} and quantum metrology \cite{na-sci-316-726}. These applications mainly rely on non-classical interference, so one constraining factor is that the photons must be in a pure quantum state \cite{ur-lasphys-15-146}. A way to generate single photons is to use pair photon sources based on parametric down-conversion or four-wave mixing (FWM), where the detection of one photon of the pair can herald the presence of the second photon \cite{so-jmo-54-467}. However, due to energy and momentum correlations, the heralded photon usually ends up in a mixed state \cite{ga-oe-15-14870}. The common way to counteract this issue is then to strongly filter the photon pair at the cost of losses, which significantly reduces the overall efficiency $\mu$ of detecting a photon. In a multi-photon experiment involving $n$ photons (such as a cascaded CNOT gate), the overall probability of detection scales as $\mu^n$ meaning that a high effective detection efficiency is essential. It has recently been suggested that naturally narrow band phase matching can be achieved in the case of a birefringent photonic crystal fibre (PCF) when the photon pairs are created in the orthogonal polarization to that of the pump light \cite{gr-pra-64-063815}. In this paper we study the tailoring of photon spectral properties generated by four-wave mixing in such birefringent PCF. The aim is to produce intrinsically narrow-band photons and hence to achieve high non-classical interference visibility without any requirement for spectral filtering.  We examine experimentally how the phase-matching of several different PCFs compares to theoretical expectations, and perform non-classical interference to measure the purity- first between photons from two separate PCFs, and then from a single PCF pumped in two directions using a Sagnac loop configuration. We then use simulations to look at how the purity of the photons' state depends upon the fibre parameters and the pump laser and compare these to experimental results.  Finally, we use the Sagnac setup to demonstrate a PCF source of polarization-entangled photon pairs, and perform quantum state tomography on the generated photons to quantify the entanglement present.

\section{Four-Wave Mixing In Birefringent Fibres}

At the quantum level the FWM process can be regarded as the virtual absorption of two pump ($p$) photons and subsequent creation of a signal ($s$) and idler ($i$) photon pair. The output state of the pair is then given by

\begin{equation}
\label{psi}
	|\Psi\rangle = \int\int d\omega_s d\omega_i F(\omega_s,\omega_i) |\omega_s\rangle|\omega_i\rangle
\end{equation}	
where the joint spectral amplitude (JSA) $F(\omega_s,\omega_i)$ is equal to the product of the pump amplitude $\alpha(\omega_s+\omega_i)$
and a phase matching function $\phi(\omega_s,\omega_i)$.\\

The process then has to satisfy energy conservation
\begin{equation}
\label{omega}
	\Delta \omega=2\omega_p-\omega_s-\omega_i=0
\end{equation}	
and a significant build up of amplitude in a length of fiber requires phase matching
\begin{equation}
\label{delta}
	\Delta k=2\frac{n_p\,\omega_p}{c}-\frac{n_s\,\omega_s}{c}-\frac{n_i\,\omega_i}{c}-2\gamma P_{p}=0
\end{equation}\noindent
where $n_{p,s,i}$ are the refractive indices of the medium at pump, signal and idler wavelengths, $P_{p}$ is the peak pump power, and $\gamma$ is the nonlinear coefficient of the fibre:
\begin{equation}
\label{gamma}
\gamma=2\pi n_{2}/\lambda_{p}A_{eff}
\end{equation}
\noindent
with $n_{2}=2\times 10^{-20}$ m$^{2}/W$ the nonlinear refractive index of silica, and $A_{eff}$ the effective cross-sectional area of the fibre mode \cite{agrawal}.

\begin{figure}[tp]
	\begin{center}
		\includegraphics[width=0.9\textwidth]{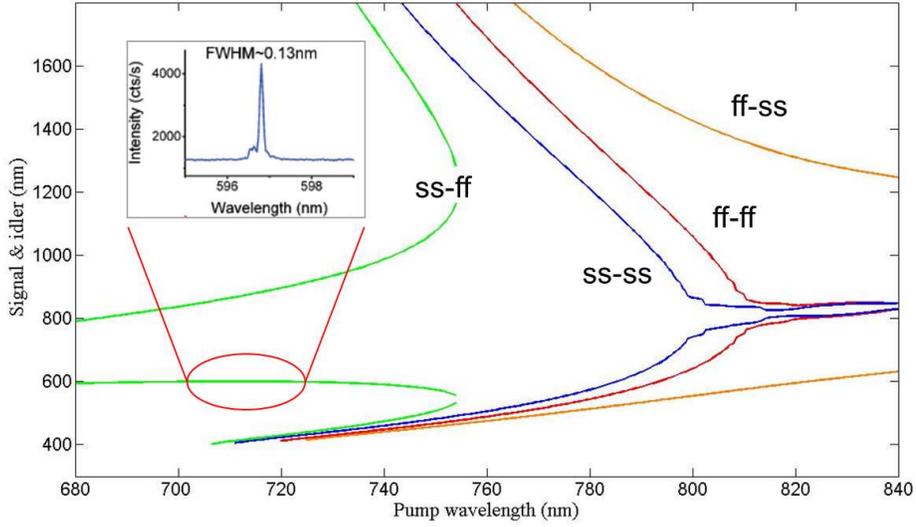}
		\caption{Signal and Idler wavelength as a function of Pump in a Birefringent fibre}
		\label{fig:Fig1}
	\end{center}
\end{figure}

For a birefringent fibre, the refractive index will be different along the slow (s) and fast (f) axes. Solving equations \ref{omega} and \ref{delta} will result in four different solutions as illustrated in Fig. \ref{fig:Fig1}. Either the pair is created along the same axis as the pump photons ($ss \rightarrow ss$ and $ff \rightarrow ff$) or the pair is created along the axis orthogonal to that of the pump ($ss \rightarrow ff$ and $ff \rightarrow ss$). In the fabrication process of the fibre, a careful choice of $n_{p}$ and $n_{i}$ can lead to a situation where for an appropriate $\lambda_p$ the group velocity of the pump is equal to that of the idler \cite{ga-oe-15-14870}. This situation corresponds to the zero slope region of the $ss \rightarrow ff$ curve in Fig. \ref{fig:Fig1} giving rise to signal photons that are intrinsically narrowband and insensitive to pump wavelength over a short range. In this case the photon-pairs are created in an uncorrelated spectral state with a JSA that is factorable: $F(\omega_s,\omega_i)=S(\omega_s).I(\omega_i)$, and the detection of one photon of the pair can then be used to herald the other in a pure quantum state.

\section{Experimental study}

\begin{figure}[h!]
\begin{center}
\includegraphics[width=0.5\textwidth]{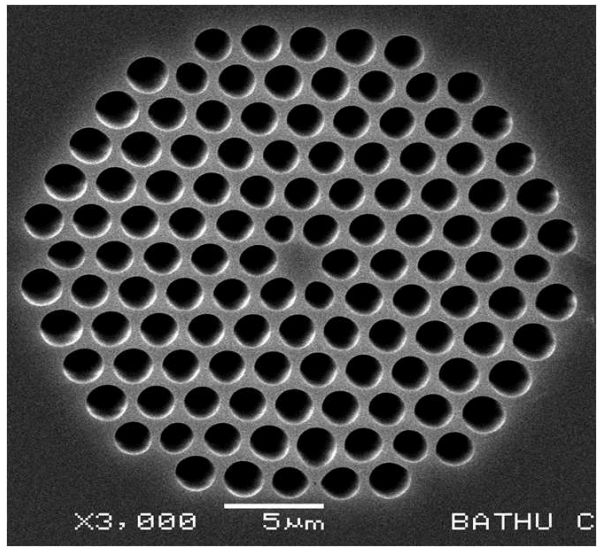}
\caption{Scanning electron microscope (SEM) image of the central region of PCF A. The birefringence required was introduced by reducing the size of two holes adjacent to the core (11 o\textquoteright clock and 5 o\textquoteright clock).}
\label{fig:SEM}
\end{center}
\end{figure}

We have fabricated a set of birefringent PCFs (-A, -B, and -C) designed to have a zero dispersion wavelength (ZDW) around $780-800\, nm$. These fibres have similar dispersion profiles, but exhibit a slightly different birefringence due to different air hole size in the cladding as can be seen if Fig. \ref{fig:SEM}. The ZDW and the birefringence were chosen to satisfy the cross polarization phase matching $ss \rightarrow ff$ in the pumping region $700-740\,nm$ accessible by our Ti:Sapphire laser. The zero slope section of the signal phase matching shown previously in Fig. \ref{fig:Fig1} was made to sit in the region $600-650\, nm$. In this manner, the matching idler photons are emitted in a range where Si APDs are still efficient ($800-870\, nm$), but are far enough from the pump wavelength to keep the Raman background low.

\begin{figure}[h!]
	\centering
		\includegraphics[width=0.70\textwidth]{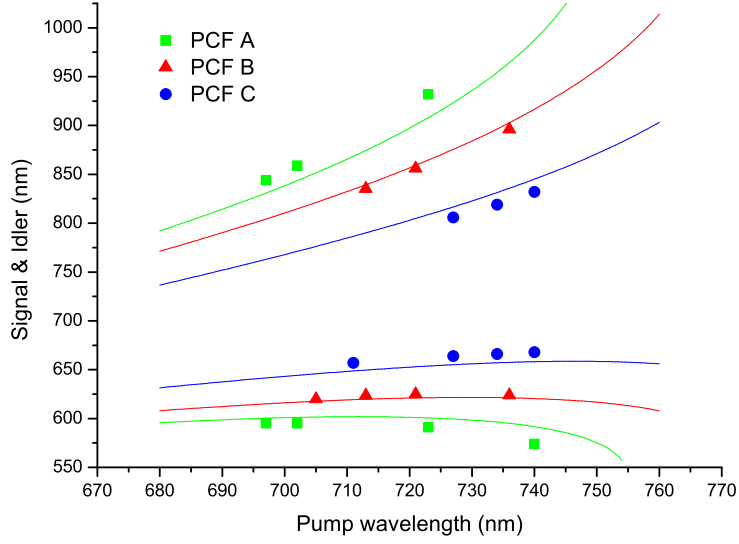}
	\caption{Theoretical phase matching curves and the corresponding experimental results in the $ss \rightarrow ff$ cross polarization phase matching case, for a set of three birefringent PCFs (-A, -B, and -C).}
	\label{fig:plot(s-i,p)}
\end{figure}

Using a plane-wave method \cite{po-optex-11-22}, we can calculate the dispersion curves and retrieve the phase matching solutions for the three fabricated PCFs. We show in Fig. \ref{fig:plot(s-i,p)} the theoretical phase matching for the $ss \rightarrow ff$ cross polarization case and the corresponding experimental results when pumping from $700\, nm$ to $740\, nm$, which are in good agreement.

\subsection{Spectral bandwidths}

\begin{figure}[h!]
	\centering
		\includegraphics[width=0.70\textwidth]{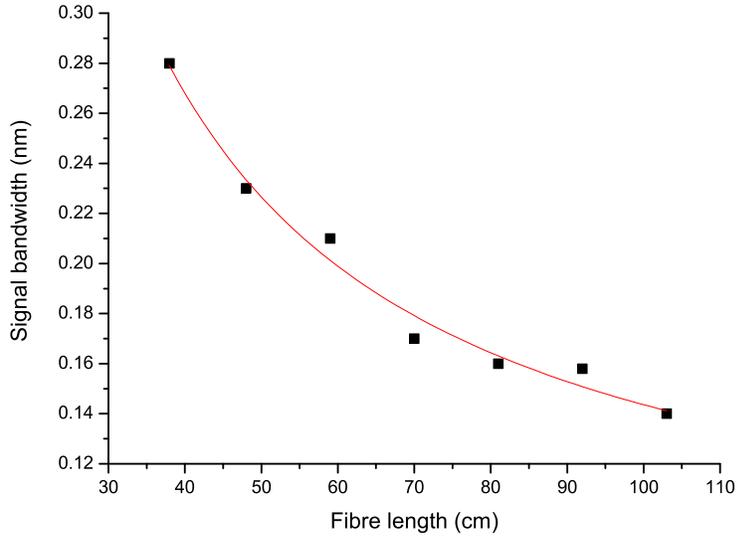}
	\caption{Spectral bandwidth of the signal photons as a function of fibre length and a 1/L fitting curve.}
	\label{fig:plot(ds,L)}
\end{figure}

We now focus on the zero slope region of the phase matching, and study the spectral bandwidth of the signal photons as a function of fibre length.
As described in \cite{al-njp-8-67}, the bandwidth equation of the generated photons is given by
\begin{equation}
\Delta\omega_{s}=\frac{2\pi c}
{\left|\mathcal{N}_s-\mathcal{N}_i\right|L}
+2\left|\frac{\mathcal{N}_i-\mathcal{N}_p}{\mathcal{N}_s-\mathcal{N}_i}\right|\Delta\omega_p\label{bdw}
\end{equation}

\noindent where $ L$ is the fibre length.

This equation is thus driven by two terms, one corresponding to the natural bandwidth of the phase matching, and a second related to the pump bandwidth broadening. The first is proportional to the inverse of the fibre length while the second is proportional to the slope of the phase matching curve. In our case, sitting on the zero slope region of the phase matching curve of the signal photons means that the signal bandwidth is insensitive to pump bandwidth, and thus only behaves as the inverse of the fibre length. This is verified by our measurement shown in Fig. \ref{fig:plot(ds,L)} where we can see the experimental data and the corresponding $1/ L$ fitting curve. This study was limited by the fact that homogeneity over the length of fibre is very difficult to control during the fabrication process and 1m of fibre is the longest usable length. As we will see in the next section, this limitation will also restrain our experimental study on purity.

\subsection{Quantum interference}

\begin{figure}[bp]
\centering
\includegraphics*[width=0.70\textwidth]{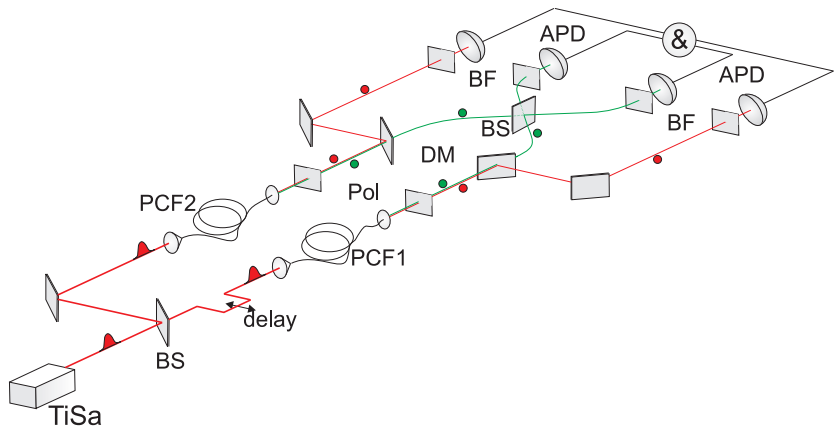}
\caption{Schematic of the experimental setup to demonstrate quantum interference between photons generated in separate fibres.}
\label{HOMsetup}
\end{figure}

In order to experimentally test the purity of our heralded single photons, we performed a Hong-Ou-Mandel (HOM) experiment \cite{ho-prl-59-2044} using heralded single photons generated from independent PCF sources. Fig. \ref{HOMsetup} shows the setup of the experiment. Amongst the three PCFs fabricated, we decided to use PCF-A which has the highest birefringence. In this manner, pumping at $705\, nm$ corresponding to the zero slope region of the signal curve, phase matching is satisfied for generation of signal and idler photons at $597\, nm$ and $860\, nm$ respectively. At these wavelengths, the detection efficiency is high ($\sim60\%$ for the signal and $\sim40\%$ for the idler) and the Raman background in the idler arm remains low. We use a mode-locked Ti:Sapphire pump laser (TiSa) emitting $100\, fs$ pulses at $705\, nm$ with a repetition rate of $80\, Mhz$ which passes through a tuneable filter allowing control of the bandwidth from $7\, nm$ ($\sim 100\, fs$) to $0.25\, nm$ ($\sim 3\, ps$). Within this full range, the signal bandwidth remains constant at $\sim 0.15\, nm$ demonstrating its independence to pump wavelength in this region. The light is then sent onto a 50:50 beam-splitter (BS) to pump two separate $40\, cm$ long pieces of PCF-A. At the output of each PCF, a polarizer (Pol) oriented with the polarization of the pairs is used to filter the remaining pump light in the orthogonal polarization, and a dichroic mirror (DM) centered at $700\, nm$ is used to split signal and idler photons into two different arms. In order to get rid of the residual pump light, two bandpass filters (BF) of $40\, nm$ and $10\, nm$ bandwidth are used in the signal and idler arms respectively. The idler photons of each source are then launched into single mode fibres and the signal photons into a single-mode 50:50 coupler with polarization controllers (PC) in one of the input arms. As the photons are well mode matched to these optical fibres and they have not undergone narrow spectral filtering we are able to achieve high overall lumped efficiencies for both sources of $\eta_{s}=0.25$ and $\eta_{i}=0.18$ for signal and idler respectively.  All four outputs are finally connected to Silicon avalanche photodiodes (APD) linked to a four-fold coincidence measurement apparatus ($\&$).\\

\begin{figure}
\centering
\includegraphics*[width=0.70\textwidth]{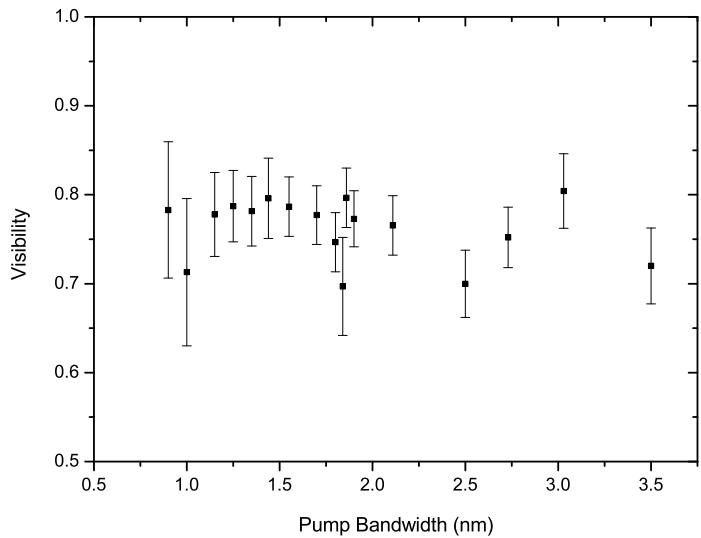}
\caption{HOM interference visibility plotted as a function of pump bandwidth.}
\label{plotVdp}
\end{figure}

\begin{figure}
\centering
\includegraphics*[width=0.70\textwidth]{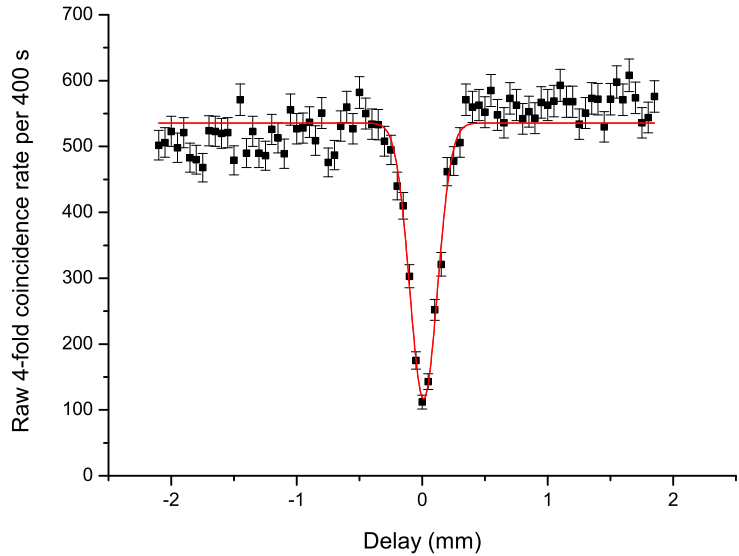}
\caption{Coincidences as a function of delay position with a pump bandwidth of $3.03\, nm$.}
\label{bestdip}
\end{figure}

A retro-reflector mounted on a micrometric translation stage in one arm of the interferometer allows us to vary the temporal overlap ($\delta t$) between the two photon pairs. When the two signal photons are made to arrive simultaneously on the beam splitter ($\delta t=0$), we can see a reduction of the four-fold coincidence count rate, which gives a measure of the purity of our heralded state. Using the tuneable filter, we have performed this experiment for different pump bandwidths. The visibility of the HOM dip as a function of pump bandwidth is plotted in Fig. \ref{plotVdp}. The visibility seems reasonably constant across the range shown with a weak peak when the pump bandwidth is $\sim 3.03\, nm$ where a $80\pm2.5\%$ visibility dip can be observed (see Fig. \ref{bestdip}). Although a clearly non-classical result, the visibility falls below the theoretical maximum of $100\%$. A possible reason for this is that the separate fibres produce signal photons with slightly different spectra due to variation in the fabrication process. To investigate this, we performed an experiment where both photons are created in the same fibre, and are thus spectrally indistinguishable.

\subsection{Quantum interference from a bi-directional source}

To test whether this limit to the visibility of the two photon interference was due to the use of two separate lengths of PCF we decided to try a method of generating two pairs of photons in one length of fibre, as was previously carried out in reference \cite{co-prl-102-123603}.  In this scheme the pump pulse is separated into two pulses that are set to counter-propagate through opposite inputs of the same fibre, but on the same principle axis.  This is achieved through a Sagnac loop configuration which is also useful in generating entangled photons.

\begin{figure}[h!]
\begin{center}
\includegraphics*[width=0.7\textwidth]{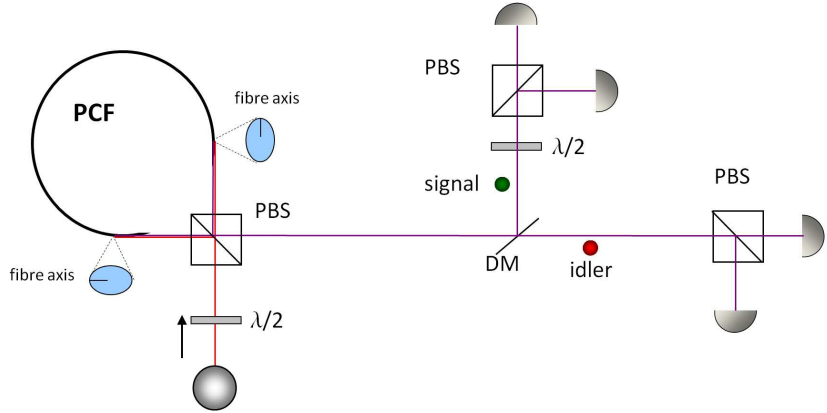}
\caption{The setup used to generate two pairs of photons from one PCF and perform a polarization based HOM dip between a photon of each pair.}
\label{fig:bidirectional_interference_setup}
\end{center}
\end{figure}

The setup for this experiment can be seen in Fig. \ref{fig:bidirectional_interference_setup}. A 20 cm length of the birefringent PCF used previously is set up on two alignment stages with the ends able to rotate about the axis of propagation, allowing the fibre to be aligned along either birefringent axis.  A polarizing beam splitter (PBS) is then placed at the point where the output modes of both ends of the PCF meet.  It can be seen at this point that if the pump pulse is sent through the PBS it can travel to either input of the PCF dependent upon it's polarization. The PCF input for pumping clockwise around the loop is rotated so that the slow axis is horizontal, while the input for pumping counter clockwise has the slow-axis orientated vertically.

Pump photons propagating in a counter clockwise direction are horizontally polarized and therefore these photons generate vertically polarized signal and idler photons due to the cross-polar phase matching scheme. As the fibre is twisted by $90^{\circ}$ the signal and idler polarizations are rotated to be horizontally polarized and exit through the PBS towards the detectors, leaving the output state as $|\psi\rangle_{out} = |H\rangle_{s}|H\rangle_{i}$. If the pump photons propagate in the clockwise direction then the same processes occur but $H \rightarrow V$ and $V \rightarrow H$.  The vertically polarized pump photons $|\psi\rangle_{in} = |V\rangle_{P_{1}}|V\rangle_{P_{2}}$ generate horizontally polarized signal and idler photons $|\psi\rangle = |H\rangle_{s}|H\rangle_{i}$ which are then transformed back to vertical polarization by the twist in the fibre so that $|\psi\rangle_{out} = |V\rangle_{s}|V\rangle_{i}$.  These photons therefore exit again through the same output of the PBS as in the counter clockwise case.  Any pump photons which do not generate signal and idler photons are rejected back towards the pump laser.

If the input power is large enough there is a finite probability of generating a pair of photons in both directions, so we have the state
\begin{equation}
\label{eqn:bidirectional_psi_out_4_photon}
|\psi\rangle_{out} = |H\rangle_{s}|H\rangle_{i}|V\rangle_{s}|V\rangle_{i}
\end{equation}

\begin{figure}[tp]
\begin{center}
\includegraphics*[width=0.9\textwidth]{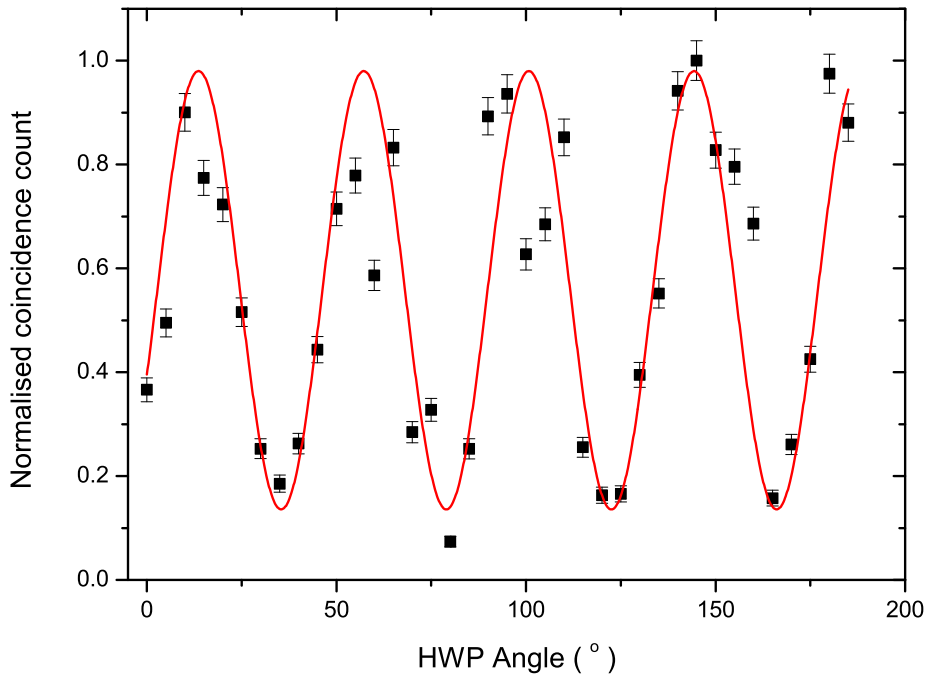}
\caption{Interference fringes in the coincidences detected as the polarization is varied with a HWP.}
\label{fig:bidirectional_interference_result}
\end{center}
\end{figure}

\noindent all in the same single spatial mode.  By inserting a dichroic mirror we can split the signal and idler photons into separate modes for heralding and analysis.  All of the photons are collected into single mode fibres with polarization controllers to correct for any rotation caused by bends in the fibre and sent to analysis benches.  The idler photons are split into two polarization modes by a PBS and then detected using silicon APD's as before, heralding the signal photons in the setup.  The signal photons are passed through a HWP in a rotation mount and then sent through another PBS, followed by detection. Here we use the HWP in the signal arm to perform a polarization analogue of the standard delay-based HOM dip, where the polarizations of the signal photons are rotated from H and V to $+45^{\circ}$ and $-45^{\circ}$ respectively, and the degree of suppression of the heralded (and therefore four-fold) coincidence detection rate across all output ports characterizes the purity and distinguishability of the incident photons \cite{co-prl-102-123603}.  With a pump bandwidth of $\sim 0.7\, nm$ and signal and idler filtering the same as for the previous HOM dip experiment, we see in Fig. \ref{fig:bidirectional_interference_result} fringes of $\sim77\%$ visibility, again below the theoretical maximum of $\sim100\%$.  To discern whether there is an inherent purity limitation with the PCF used we construct a model to compare to experiment.

\section{Schmidt number optimization}

The purity of the heralded state is determined by the degree of factorability of the JSA \cite{mo-njp-10-093011} and varies with pump wavelength, fibre length, pump bandwidth and fibre geometry (by which the group velocity for the signal, idler and pump wavelengths can be controlled). We have developed a piece of software to optimise these variables for maximum purity. A general JSA can be decomposed into its Schmidt modes

\begin{equation}
F(\omega_s, \omega_i)=\sum_{j} \sqrt{\lambda_j}f_j(\omega_s)g_j(\omega_i)
\end{equation}

\noindent where $f_j(\omega_s)g_j(\omega_i)$ are an orthonormal set of completely factorable modes. To quantify the factorability, we use the singular value decomposition (SVD), which is the matrix analogue of the Schmidt decomposition. We define the matrix representation of the JSA as $M=A.\Phi$, where $A$ is the matrix corresponding to the pump amplitude and $\Phi$ to the phase matching function. The SVD is then a factorisation of $M$ into three matrices, which will be called I,S,D: the columns of I represent the part of the Schmidt modes relating to the idler while the rows of S represent the modes of the signal, and D is a diagonal matrix linking the two. The elements of the diagonal matrix contain the amplitudes $\sqrt{\lambda_j}$ and are in descending order of magnitude.

We obtain D from the SVD and calculate the Schmidt magnitudes $\lambda_j$, with the following normalisation:
\begin{equation}
\sum_{j} \lambda_j=1
\end{equation}
\indent A completely pure state will be represented by $\lambda_{j=1}=1$ and $\lambda_{j\neq1}=0$. The Schmidt number $K$ is then defined as
\begin{equation}
K=\frac{1}{\sum_{j} \lambda^2_j}
\end{equation}
\indent For a pure state, $K$ will equal 1. For a completely entangled state, there will be an infinite number of Schmidt magnitudes $\lambda$, each approaching zero. The sum of the magnitudes will therefore approach zero with the result that $K\rightarrow\infty$. We have previously seen that the JSA is defined as $F(\omega_s,\omega_i)=\alpha(\omega_s+\omega_i)\phi(\omega_s,\omega_i)$, where

\begin{equation}
\label{alpha}
	\alpha(\omega_s+\omega_i)=e^{-\frac{(\Delta\omega_s+\Delta\omega_i)^2\sigma^2}{2}}
\end{equation}

\noindent and

\begin{equation}
\label{phi}
	\phi(\omega_s,\omega_i)=\left(e^{\frac{iKL}{2}}\right)\left(sinc(\frac{\Delta k L}{2})\right)
\end{equation}

\indent Following the same steps described in \cite{al-njp-8-67}, we develop the phase-matching equation around the phasematched frequencies using a first order Taylor series in order to express $\Delta k$ in terms of the group index.

\begin{eqnarray}
\Delta k &=& 2k_p-k_s-k_i-2\gamma P \\
&=& 2\left.\frac{\partial k_p}{\partial\omega}\right|_{\omega_p}\Delta\omega_{p}-\left.\frac{\partial k_s}{\partial\omega}\right|_{\omega_s}\Delta\omega_{k_s}-\left.\frac{\partial k_i}{\partial\omega}\right|_{\omega_i}\Delta\omega_{k_i} \nonumber
\end{eqnarray}

\indent With the help of energy conservation, we can substitute $\Delta\omega_p$ and find:

\begin{equation}\label{dk}
	\Delta k=\frac{\left(\mathcal{N}_s-\mathcal{N}_p\right)}{c}\Delta\omega_{k_s}+\frac{\left(\mathcal{N}_i-\mathcal{N}_p\right)}{c}\Delta\omega_{k_i}
\end{equation}

\noindent where $\mathcal{N}_l=\left[\omega_l\left.\frac{\partial n_l}{\partial\omega}\right|_{\omega_l}+n_l\right]$ is the group index for mode l. The JSA, and therefore $M$ can now be expressed as a function of $\Delta\omega_s$ and $\Delta\omega_i$ only. Using the Sellmeier equation \cite{hecht}, we can then generate the group index for a given fibre geometry and study the factorability of the JSA for a given phase-matching function, pump bandwidth and fibre length.\\

\begin{figure}[tp]
\centering
\mbox{\subfigure[ ]{\includegraphics[width=0.46\textwidth]{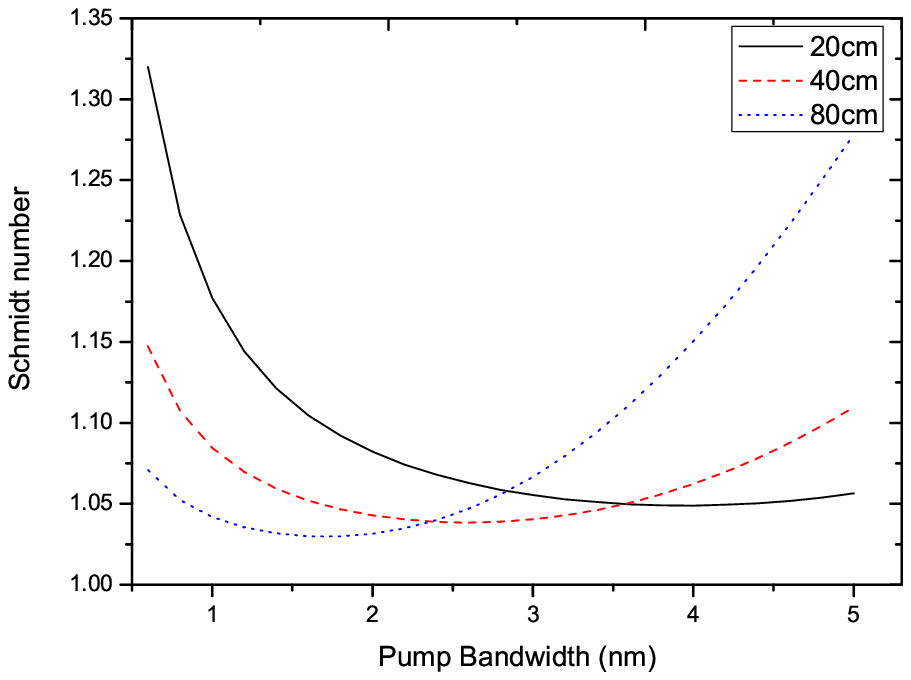}\label{plot(k,lp,L)}}\quad
\subfigure[ ]{\includegraphics[width=0.5\textwidth]{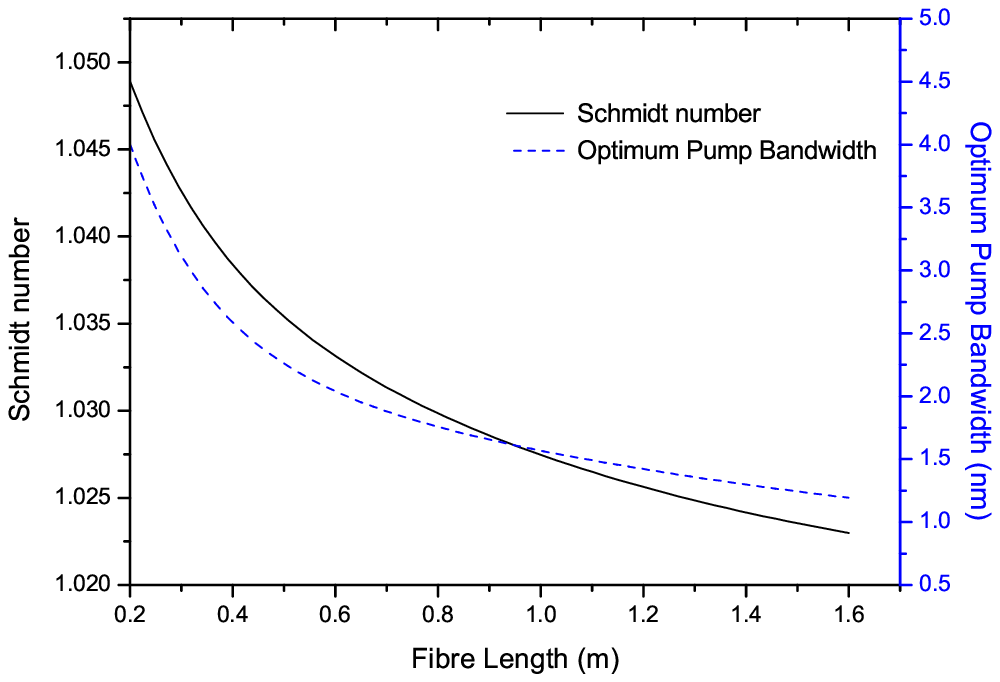}\label{plot(k,lp,L)2} }}
\caption{(a) Schmidt number of the heralded state as a function of pump bandwidths for three different fibre lengths: $20$, $40$, and $80\, cm$. (b) Minimum Schmidt number and the corresponding pump bandwidth as a function of fibre length.}
\end{figure}

\begin{figure}[tp]
\centering
\mbox{\subfigure[ ]{\includegraphics[width=0.5\textwidth]{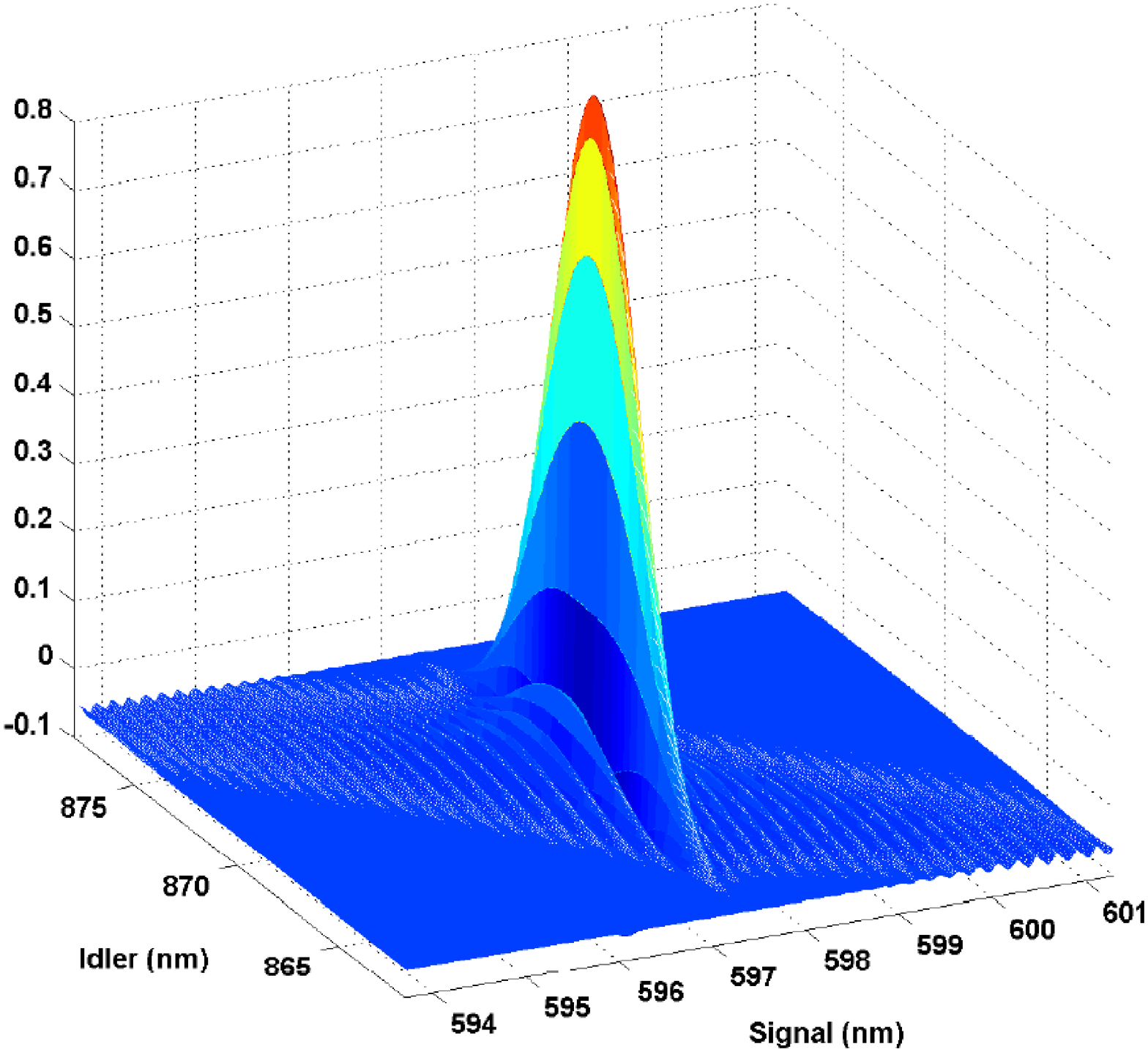}\label{plot(F)1}}\quad
\subfigure[ ]{\includegraphics[width=0.5\textwidth]{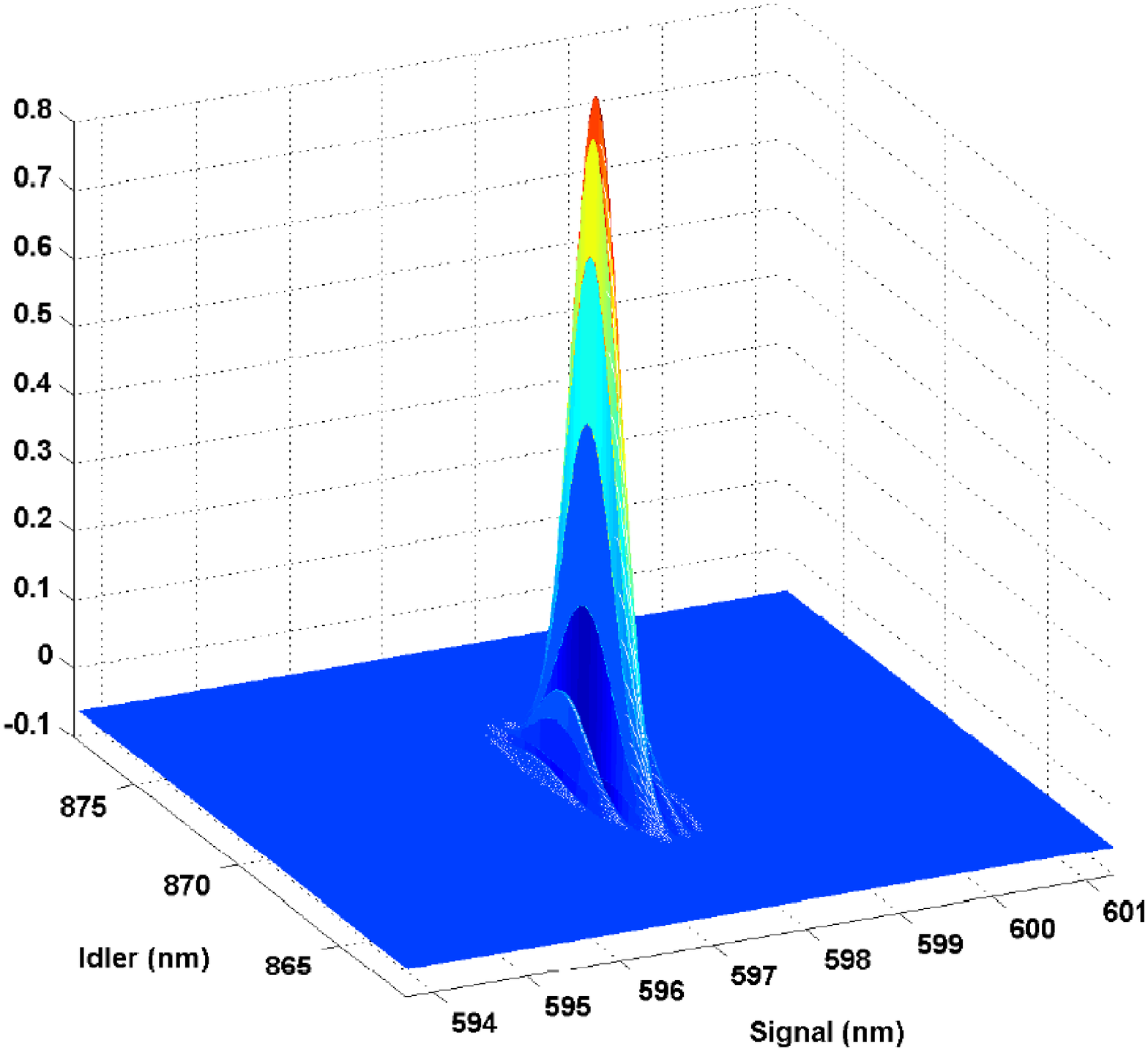}\label{plot(F)2} }}
\caption{JSA of the created photon pair given by the product of the phase matching function and the envelope of the pump intensity (a) Unfiltered case (b) Where large bandpass filters ($\Delta \lambda_{filter} \gg \Delta \lambda_{signal/idler}$) have been applied to both photons.}
\end{figure}

We first look at the purity of the heralded state as a function of pump bandwidths for three different fibre lengths: $20$, $40$, and $80\, cm$. The result of the simulation is shown in Fig. \ref{plot(k,lp,L)}. As we can see, each fibre length has an optimum pump bandwidth corresponding to a minimum of the Schmidt number. We then plot in Fig. \ref{plot(k,lp,L)2} the minimum Schmidt number accessible and the corresponding optimum pump bandwidth as a function of fibre length. This shows that the longer the fibre the better purity (lower Schmidt number) we can get. It is also interesting to note that the optimum pump bandwidth decreases when the fibre length increases. One explanation is that when the pump bandwidth is wide, both signal and idler are limited in width by the phase-matching function, while when it is narrow, both are limited by energy conservation. At the optimum bandwidth the signal is mainly limited by the phase-matching and the idler mainly by energy-conservation, so their state is more factorable. As the fibre length is increased, the phase-matching $sinc$ function decreases in width (equation \ref{phi}), and so the pump bandwidth can also be decreased without reaching the region where it limits the signal bandwidth.

\begin{figure}[h!]
\centering
\includegraphics[width=0.5\textwidth]{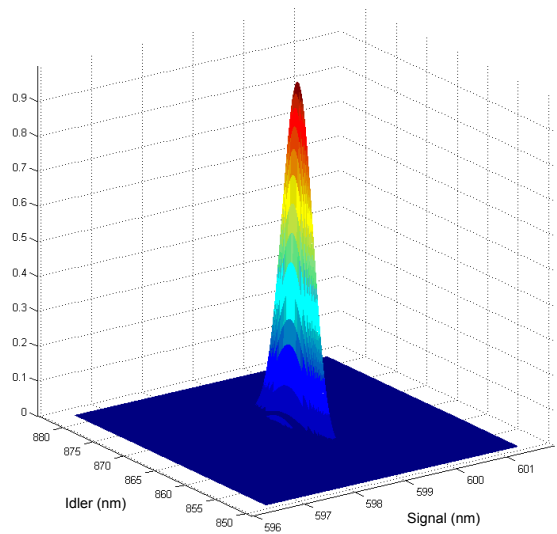}
\caption{Unfiltered Joint Spectral Intensity for a length of $40\, cm$ and pump bandwidth $2.6\, nm$.}
\label{JSI}
\end{figure}

\begin{figure}[h!]
\centering
\mbox{
\subfigure[ ]{\includegraphics[width=3in]{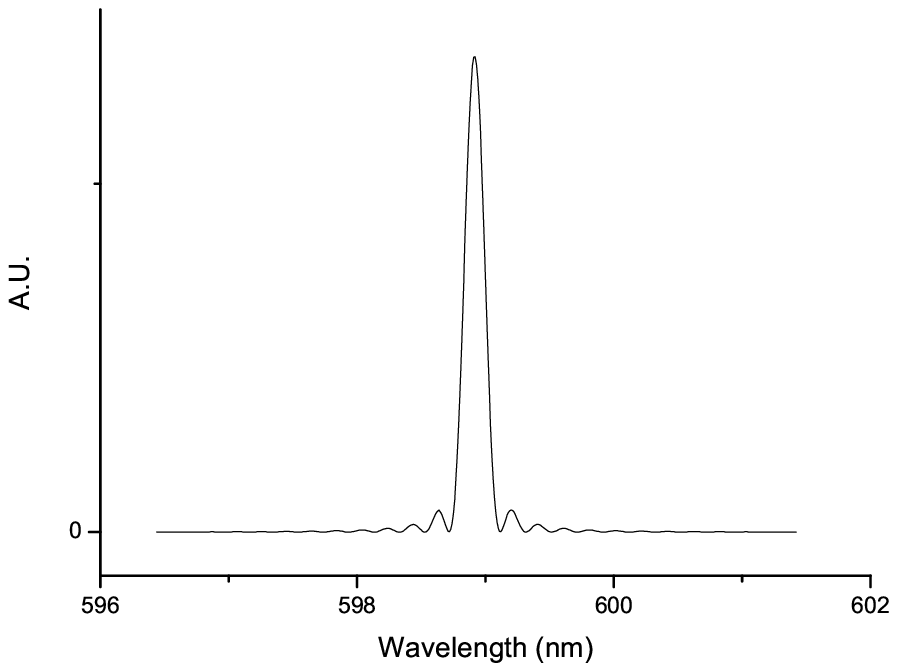}}\quad
\subfigure[ ]{\includegraphics[width=3in]{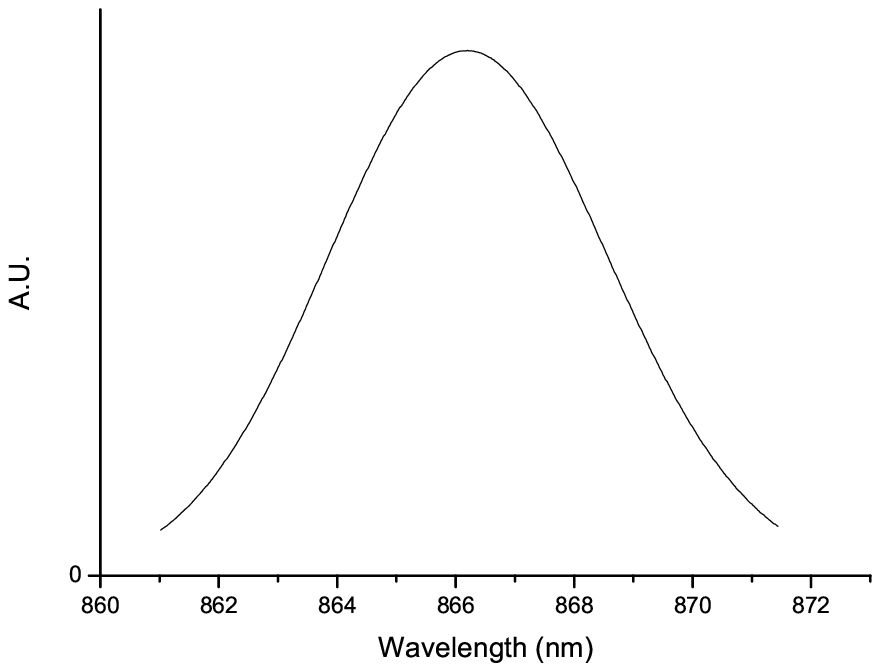}}}
\caption{Spectral intensities projected from the JSI for (a) the signal (b) the idler.}
\label{spectraplots}
\end{figure}

Fig. \ref{plot(k,lp,L)2} shows that even in the case where the pump bandwidth has been optimised for a given fibre length, the Schmidt number still does not reach the optimum value of 1. Looking at the JSA for an example of fibre length of $40\, cm$ (Fig. \ref{plot(F)1}), we can see that there are still some correlations between the signal and idler frequencies coming from the residual ripples of the phase matching $sinc$ function. These ripples will appear small in the measurable spectrum, shown as a Joint Spectral Intensity (JSI) in Fig. \ref{JSI} and spectral intensities of the signal and idler in Fig. \ref{spectraplots}, but the resulting correlations in the JSA will increase the Schmidt number. A way to enhance the purity of the heralded photon is then to remove these ripples using bandpass filters whose bandwidths can be much greater than the full-width half maximum (FWHM) of the generated photons. This will result in an improvement of the purity without involving significant losses in photon collection efficiencies. Fig. \ref{plot(F)2} shows the same fibre length as in Fig. \ref{plot(F)1} where bandpass filters were applied to both signal and idler photon spectra and the pump bandwidth has been optimized to this new configuration. As a result, we can see that most of the ripples have disappeared and the calculated Schmidt number has decreased from 1.04 without filters to 1.01 in the filtered case. It is important to note that the photons can still be considered as unfiltered since the filters used satisfy the following condition: $\Delta \lambda_{filter}\gg \Delta \lambda_{signal/idler}$.

\begin{figure}[h!]
\centering
\includegraphics[width=0.8\textwidth]{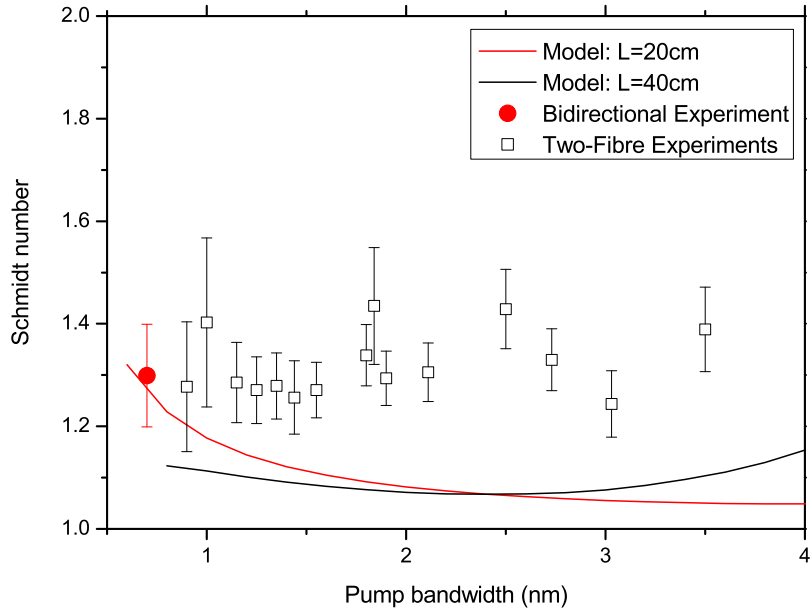}
\caption{Comparison of experimental and theoretical Schmidt number for a variety of fibre lengths.  Black squares show two fibre experiment points while red circles show the bidirectional results.}
\label{compareexperimentplot}
\end{figure}

We expect the Schmidt number to be related to the HOM dip visibility by $V=1/K$.  To determine whether the HOM dip visibilities seen experimentally were limited by correlations in the JSA we can compare the Schmidt numbers derived from the experimental results with those obtained from the simulation for a range of pump bandwidths and the experimental conditions used.  We show in Fig. \ref{compareexperimentplot} that there is an approximately constant offset between the experimental and theoretical results for the two fibre case, with the experimental Schmidt numbers never reaching the expected low values of the simulation, suggesting there is another reason for the lack of photon purity.  To test whether this difference arises from the use of two separate pieces of fibre we can use the visibility of the fringes from the bidirectional case to estimate a Schmidt number for this fibre length and pump bandwidth, and also compare this to the simulation.  In doing so we see in Fig. \ref{compareexperimentplot} that there is no offset present and the theoretical Schmidt number lies within the error of the experimental point, showing that the lack of visibility could be entirely explained by correlations between signal and idler photons being present in the JSA.  This would confirm that the extra loss of visibility in the two fibre experiment is explained by differences between the two lengths of fibre introducing some cross-source distinguishability.

We can also note that if the bidirectional source was pumped with a larger pump bandwidth of $\sim 4\, nm$ we can expect a fringe visibility of over $95\%$ and a Schmidt number below 1.05.  Further experimental investigation is being carried out to confirm this result. We note also that the pump filtering method produces top hat pump envelope shapes which increase the Schmidt number from that which is achieved through a Gaussian pump envelope of similar bandwidth.

\section{Entangled photon source}

The Sagnac loop configuration used above has another intrinsic use when used at lower powers, namely the generation of entanglement.  The use of entangled photons is central to many quantum communication and computation schemes \cite{gi-natphot-1-165, kn-nat-409-46, br-prl-95-010501} and plays a vital role in the development of quantum repeaters \cite{br-prl-81-5932} and quantum memories \cite{he-nat-465-1052}.  The Sagnac loop method has been well demonstrated previously for degenerate and non-degenerate photons, both in fibre \cite{li-prl-94-053601, fu-prl-99-120501} and using nonlinear crystals \cite{fe-opx-15-23}, however all implementations required filtering and therefore suffer from low counting rates.  Here we show a demonstration of a source of entangled photons which are in an intrinsically pure state from a PCF in a Sagnac loop, therefore requiring no narrow spectral filtering.

\begin{figure}[h]
\begin{center}
\includegraphics*[width=0.8\textwidth]{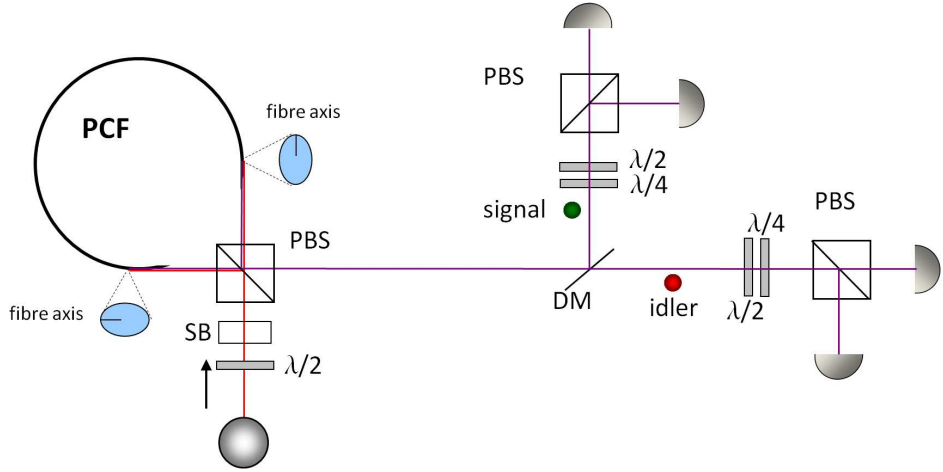}
\caption{The setup used to generate pairs of entangled photons from a PCF in a Sagnac loop.}
\label{fig:entanglement_setup}
\end{center}
\end{figure}

The setup for this experiment is very similar to the bi-directional interference experiment with extra wave plates for analysis and a Soleil-Babinet compensator added, as seen in Fig. \ref{fig:entanglement_setup}.  A Ti:Sa laser emits $100\, fs$ laser pulses ($\sim 8 nm$ broad) that are then filtered as before, using two diffraction gratings and a slit, down to around $1\, nm$.  This is then passed through a HWP which allows us to set a linear superposition of horizontal and vertical polarization followed by a Soleil-Babinet compensator consisting of two birefringent crystal wedges that can slide across one another, thereby imparting a variable phase $\phi$ between horizontal and vertical components.  With this combination any pump state $|\psi\rangle_{in}$ can be set for the input of the system.

\begin{equation}
\label{eqn:entangled_psi_in}
|\psi\rangle_{in} = \alpha |\gamma\rangle_{H} + \beta e^{i\phi}|\gamma\rangle_{V}
\end{equation}

Here we use $\gamma$ to signify a coherent state of many photons, as are present in the pump pulses.  This pulse then enters a PBS which transmits horizontally, and reflects vertically, polarized photons.  When two horizontally polarized photons travel in the counterclockwise direction around the $20\, cm$ loop of PCF they can generate vertically polarized signal and idler photons through FWM in the cross-polar regime.  These are then converted back to horizontal polarization by a $90^{\circ}$ twist in the fibre about the axis of propagation, meaning that $|H\rangle_{P_{1}}|H\rangle_{P_{2}} \rightarrow |H\rangle_{s}|H\rangle_{i}$. When two vertically polarized pump photons travel in the clockwise direction they generate horizontally polarized signal and idler photons that are then twisted back to vertical so we have $|V\rangle_{P_{1}}|V\rangle_{P_{2}} \rightarrow |V\rangle_{s}|V\rangle_{i}$.  The twist in the fibre also ensures that photons propagating in both directions are launched onto the same axis of the PCF, so they see the same phase matching conditions.  If we then take into account the $\alpha$ and $\beta$ weightings set by the input HWP we have the following at detection

\begin{equation}
\label{eqn:entangled_general_out}
|\psi\rangle_{out} = \alpha |H\rangle_{s}|H\rangle_{i} + \beta e^{2i\phi}|V\rangle_{s}|V\rangle_{i}
\end{equation}

As the pump is orthogonally polarized to the signal and idler photons it is automatically rejected back towards the laser, reducing background noise at the output port.  We can see from equation \ref{eqn:entangled_general_out} that if we set $\alpha = \beta = \frac{1}{\sqrt{2}}$ and set $\phi = 0$ then we generate the maximally entangled $\left|\Phi^{+}\right\rangle$ Bell state.  It is worth noting that this is an inherently stable setup in that the output state is not affected by temperature fluctuations or vibrations as the relative phase between photons generated in different directions is automatically compensated \cite{jeremiethesis}.  The output state does not depend upon the length of the fibre or the point in the fibre where the photons are generated, only on the setting of the input state, making this a very versatile and useful source for quantum information applications.  At the output of the PBS the signal and idler photons are split to two paths by a dichroic mirror and then collected into optical fibres, each with a polarization controller to compensate for any stress induced polarization rotations, and then sent to analysis benches containing a QWP, HWP and PBS. The photons are then launched into multimode fibre and sent to silicon APDs on the transmitted (H) output.

\subsection{A test of entanglement}

An initial test of entanglement is a measurement of coincidence fringes. The analysis state of one photon of the entangled pair is set on one of two non-orthogonal bases, typically the computational $\{ H / V \}$ and diagonal $\{ D / A\}$ bases. The other photon is rotated continuously through the basis states and the coincidence rate between the detection of both photons is recorded. A second fringe is then measured for the other (non-orthogonal) basis. The classical $\{ H / V \}$ basis measurement should yield a coincidence fringe with unit visibility in the absence of all background noise, whilst the $\{ D / A\}$ basis measurement should be flat, or have zero visibility, for no entanglement and have unit visibility if the photons are maximally entangled.  If the visibility in this basis is greater than $71\%$ then the state is entangled enough to violate a Bell inequality \cite{be-phys-1964, cl-prl-23-880}, making this the lower bound on proving that the photons are entangled.

\begin{figure}[h!]
\begin{center}
\includegraphics*[width=0.8\textwidth]{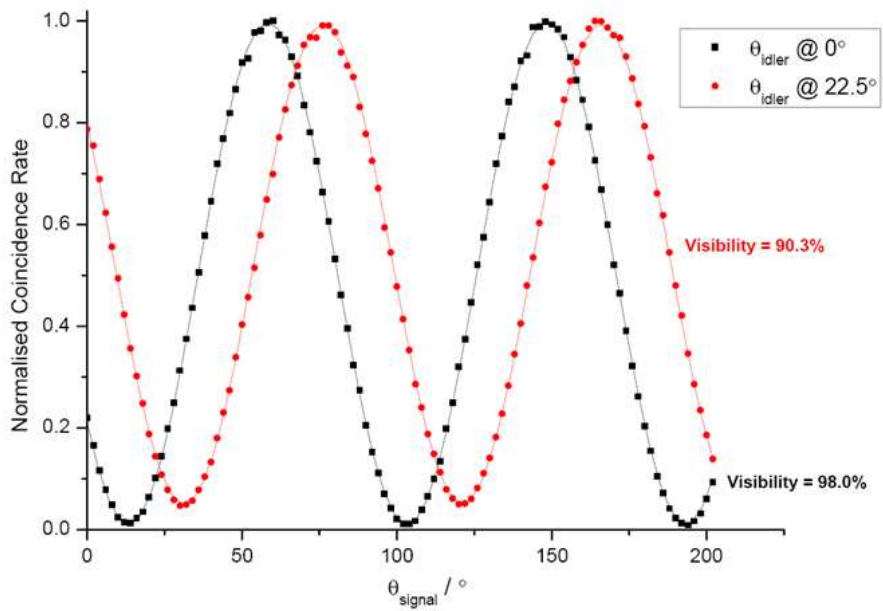}
\caption{Coincidence count fringes taken with the idler photon fixed at $0^{\circ}$ in black and $22.5^{\circ}$ in red, showing $98\%$ and $90\%$ visibility respectively.}
\label{fig:entanglement_fringes}
\end{center}
\end{figure}

For this test of entanglement the QWPs can be removed from the analysis benches for the signal and idler photons as we are only concerned with linear polarizations.  We choose the idler photon to have the fixed polarization and set the HWP in this arm to be $0^{\circ}$, or $|H\rangle$, for the $\{ H / V \}$ basis and $22.5^{\circ}$, or $|D \rangle$, for the $\{ D / A \}$ basis.  For these two positions we rotate the HWP in the signal analysis bench whilst recording single APD counts and the coincidence between the signal and idler counts.  The result in Fig. \ref{fig:entanglement_fringes} shows a visibility of $98\%$ in the $\{ H / V \}$ basis and $90\%$ in the $\{ D / A \}$ basis, clearly showing a high level of entanglement.

The level of entanglement could be reduced by reflections in the fibre \cite{jeremiethesis}, excess background noise from residual pump light, Raman photons, or most likely from an imbalanced collection of photons from one direction of propagation to the other.  This is difficult to correct for as when changing the output of one end of the loop you are changing the input coupling, and vice versa, which are often countering each other in quality.  This is, however, not a fundamental limit to the amount of entanglement possible with such a setup, which should theoretically produce maximally entangled pairs.

\subsection{Quantum state tomography}

\begin{table}[bp]
\centering
\label{tab:tomography}
\caption{Wave plate settings for two photon state tomography \cite{ja-pra-64-052312}.}
\begin{tabular}{| c |@{\extracolsep{0.7cm}} c | c | c | c | c | }
\hline	
\hline		
   $\nu$ & State & $HWP_{s}$ & $QWP_{s}$ & $HWP_{i}$ & $QWP_{i}$ \\
\hline
  1  & $|HH\rangle$ & 0 & 0 & 0 & 0 \\
  2  & $|HV\rangle$ & 0 & 0 & 45 & 0 \\
  3  & $|VV\rangle$ & 45 & 0 & 45 & 0 \\
  4  & $|VH\rangle$ & 45 & 0 & 0 & 0 \\
  5  & $|RH\rangle$ & 22.5 & 0 & 0 & 0 \\
  6  & $|RV\rangle$ & 22.5 & 0 & 45 & 0 \\
  7  & $|DV\rangle$ & -22.5 & -45 & 45 & 0 \\
  8  & $|DH\rangle$ & -22.5 & -45 & 0 & 0 \\
  9  & $|DR\rangle$ & -22.5 & -45 & 22.5 & 0 \\
  10 & $|DD\rangle$ & -22.5 & -45 & -22.5 & -45 \\
  11 & $|RD\rangle$ & 22.5 & 0 & -22.5 & -45 \\
  12 & $|HD\rangle$ & 0 & 0 & -22.5 & -45 \\
  13 & $|VD\rangle$ & 45 & 0 & -22.5 & -45 \\
  14 & $|VL\rangle$ & 45 & 0 & 22.5 & 90 \\
  15 & $|HL\rangle$ & 0 & 0 & 22.5 & 90 \\
  16 & $|RL\rangle$ & 22.5 & 0 & 22.5 & 90 \\
\hline
\end{tabular}
\end{table}

A further and more in depth analysis of the entangled state can be carried out to reconstruct the density matrix of the photon pair, known as quantum state tomography or QST \cite{ja-pra-64-052312}.  To perform QST one must take many copies of the state to be analysed and perform a tomographically complete set of measurements, where the set of measurement operators define a basis in the Hilbert space of the two photon state.  In this case that complete set of measurements is 16 polarization settings for the HWP's and QWP's in the signal and idler analysis benches. It comprises combinations of analysing $|H\rangle$, $|V\rangle$, $|D\rangle = \frac{1}{\sqrt{2}}(|H\rangle+|V\rangle)$, $|R\rangle=\frac{1}{\sqrt{2}}(|H\rangle-i|V\rangle)$ and $|L\rangle=\frac{1}{\sqrt{2}}(|H\rangle+i|V\rangle)$ states as shown in Table 1.

\begin{figure}[tp]
\begin{center}
\includegraphics*[width=0.9\textwidth]{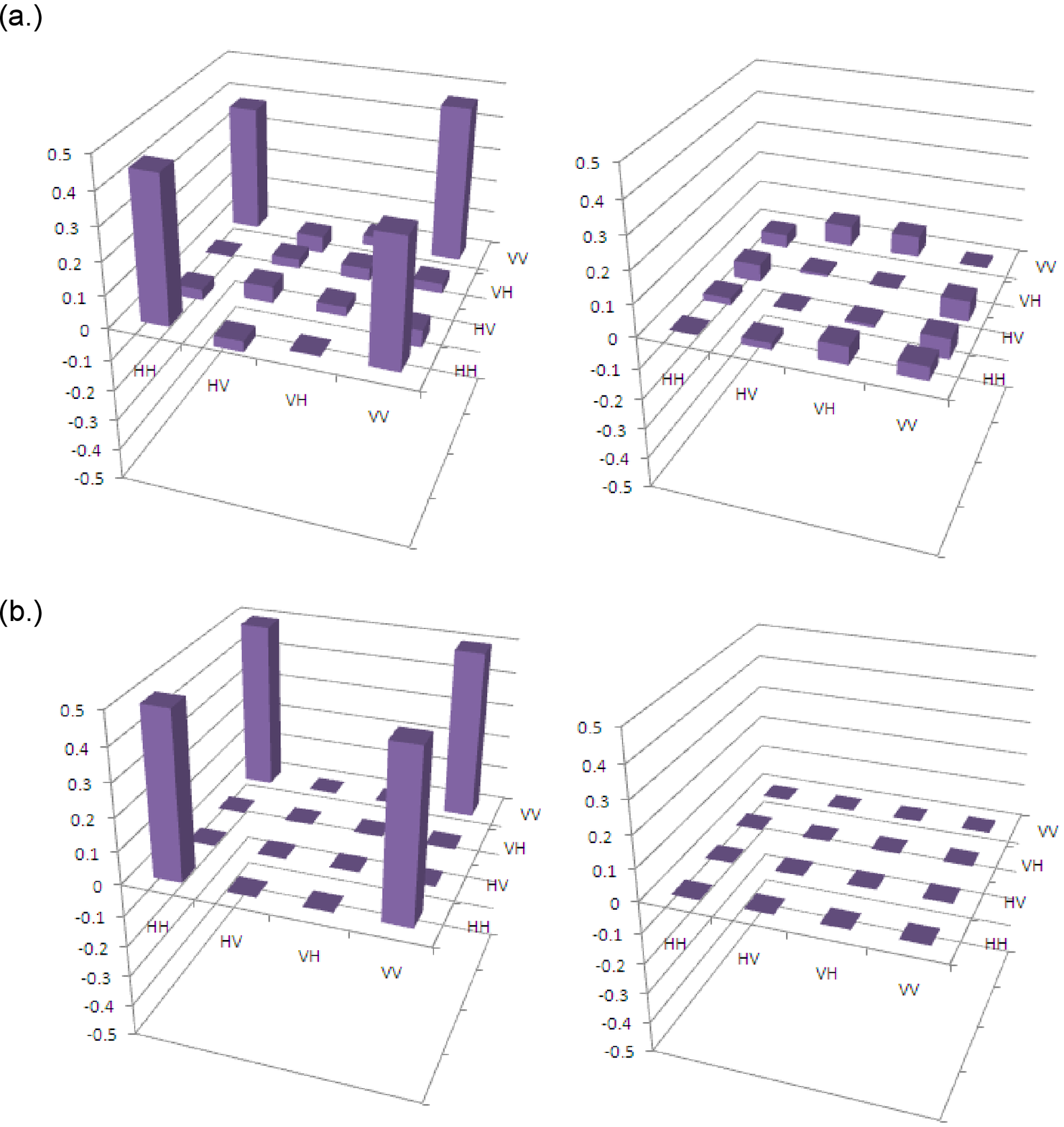}
\caption{Tomography of the generated entangled state showing a.) the reconstructed real and imaginary parts of the density matrix and b.) the real and imaginary parts of the ideal $\Phi^{+}$ state.  The experimental result shows a fidelity of $84\%$ to the $\left|\Phi^{+}\right\rangle$ state.}
\label{fig:entanglement_tomography}
\end{center}
\end{figure}

Taking the 16 counts obtained for the different measurement settings we can perform a linear tomographic reconstruction of the state.  This does, however, sometimes recreate states that are non-physical as they have negative eigenvalues and the trace of the square of the density matrix becomes greater than the allowed value of 1.  This means the density matrix $\hat{\rho}$ is not positive semi-definite as required.  To avoid this problem we employ a technique, described in \cite{ja-pra-64-052312} and \cite{ba-pra-61-010304}, known as maximum likelihood estimation.  This approach estimates the density matrix by first generating a formula for a physical density matrix and then using a likelihood function which quantifies how closely the estimated state would generate coincidences compared with the experimental data.  This process is then repeated and, using standard numerical optimization techniques, the optimum physical density matrix is found.  This process can also be improved by the use of convex optimization techniques \cite{stephaneproject}. Upon performing this calculation we find the closest density matrix to be


\begin{equation}
\label{eqn:calculated_density_matrix}
\hat{\rho}_{exp} =
\begin{pmatrix}
   0.446 & -0.033 + 0.021i & 0.002 + 0.055i & 0.382 + 0.042i \\
  -0.033 - 0.021i &  0.049 & 0.029 - 0.008i & -0.053 - 0.064i \\
   0.002 - 0.055i & 0.029 + 0.008i & 0.037  &  0.029 - 0.065i \\
   0.382 - 0.042i & -0.053 + 0.064i & 0.029 + 0.065i & 0.467 \\
\end{pmatrix}
\end{equation}

\noindent which can be seen graphically in Fig. \ref{fig:entanglement_tomography} split into it's real and imaginary parts.  Also in Fig. \ref{fig:entanglement_tomography} are the real and imaginary parts of the ideal $|\Phi\rangle^{+}$ state, which when compared to our experimentally generated state shows a fidelity of $83.93\%\pm0.08\%$.

We can also quantify the amount of entanglement generated by looking at the Tangle, $\mathcal{T}=0.5495\pm0.0027$ which is greater than $0.5$ giving a clear sign that entanglement is generated \cite{ja-pra-64-052312}.  The Concurrence $\mathcal{C}=0.7413\pm0.018$ and Linear Entropy $\mathcal{L}=0.3306\pm0.0013$ also show a degree of entanglement but are clearly not the optimum values of $\mathcal{C}=1$ and $\mathcal{L}=0$.  As mentioned before, this non-maximal entanglement is probably due to the $|H\rangle$ and $|V\rangle$ modes not being perfectly overlapped. Nevertheless, this amount of entanglement is already at a useful level, and may be improved with careful alignment.

\section{Applications and Future Work}

Many further experiments are possible with this source. The bi-directional source set-up, for instance, can be extended to 2 photon NOON-state metrology, to show phase measurements with higher precision than allowed by the standard quantum limit, at very low intensities. The idler photons are used to herald the signal photons, which are injected into a displaced Sagnac interferometer, so that when the phase in one arm of the interferometer is varied, the probability of a coincidence across both outputs shows super-resolution of fringes with a spacing of $312\, nm$: half the wavelength of the signal, and less than half that of the pump. This compares favourably to similar experiments using spontaneous parametric down-conversion sources, where the fringe spacing is equal to that of the pump \cite{ra-prl-65-1348}.

By using one entangled source and one single direction correlated source, a teleportation scheme \cite{bo-nt-390-575} between photons of different wavelengths can be demonstrated. Here, the signal photon from the correlated source heralds the presence of the idler, which can be polarization-encoded to form a qubit using wave plates. When this photon and the idler from the entangled source are mixed on a 50:50 beamsplitter and a coincidence is detected at the two outputs, the two photons have been projected onto a Bell state, and the qubit encoded previously is teleported onto the state of the remaining signal photon from the entangled source following a polarization rotation of $45^{\circ}$. Teleportation has applications in quantum communications, for instance quantum key distribution \cite{gi-natphot-1-165}, and in linear optical quantum computing \cite{kn-nat-409-46}, where the success probability of a non-deterministic gate can be improved to an arbitrary level with the addition of teleportation and a resource of entangled ancilla photons.

Similarly using one entangled source and a correlated source, a scalable version of a quantum CNOT gate can be performed \cite{pi-pra-64-062311}, between a control and target qubit encoded in photons of different wavelengths. The signal and idler photons from the correlated source are used as the target and control qubits respectively, and an input state is encoded in their polarization using wave plate rotations. The control photon is mixed on a polarizing beamsplitter (PBS 1) with the idler from the entangled pair, and the target is mixed on a separate polarizing beamsplitter rotated by $45^{\circ}$ (PBS 2) with the signal from the entangled pair. A successful CNOT operation has taken place if a diagonally polarized photon is detected at one output of PBS 1 in coincidence with a horizontally polarized photon at one output of PBS 2. The final state control and target photons are left at the remaining two outputs. The fact that a successful operation is heralded by ancilla photons, rather than relying on detecting the control and target as in \cite{cl-pra-79-030303, pi-pra-68-032316, ob-nat-426-264, ob-prl-93-080502, ra-pra-65-062324, la-prl-95-210504, ki-prl-95-210505}, means this CNOT is scalable and could be used as part of a more complex operation.

Another application is in the generation of larger entangled states such as cluster states. In the cluster-state, or one-way, model of quantum computation, a multipartite entangled state is used as a resource for running a quantum algorithm represented by a series of single-qubit measurements\cite{ra-prl-86-5188}. Using two entangled sources and a polarizing beam splitter acting as a fusion gate we can generate a four photon cluster state and demonstrate quantum gates in the one-way model. Fusion gates are a resource efficient way of building up cluster states\cite{br-prl-95-010501}, and we can extend this to larger states by adding more entangled pairs and fusion gates, or by encoding more qubits in other degrees of freedom, such as path or time-bin encoding.

In conclusion, we have demonstrated a fibre-based source of photon pairs which avoids two significant limitations on the efficiency of conventional downconversion sources: the flat phasematching curve means the photons are created in an intrinsically pure state without the need for spectral filtering, and the photons are created in a single-mode, so we achieve good spatial modematching to fibre. We have also shown high visibility non-classical interference between photons from the same and separate sources, and generated entangled pairs with a good fidelity to a maximally entangled Bell state. This makes the source suitable for many quantum information applications and future experiments.\\

\section{References}

\bibliography{index}

\begin{thebibliography}{10}

\bibitem{gi-natphot-1-165}
N.~{Gisin} and R.~{Thew}.
\newblock {Quantum communication}.
\newblock {\em Nature Photonics}, 1:165--171, 2007.

\bibitem{kn-nat-409-46}
E~Knill, R~Laflamme, and G~J Milburn.
\newblock A scheme for efficient quantum computation with linear optics.
\newblock {\em Nature}, 409:46--52, 2001.

\bibitem{na-sci-316-726}
Tomohisa Nagata, Ryo Okamoto, Jeremy~L. O'Brien, Keiji Sasaki, and Shigeki
  Takeuchi.
\newblock {Beating the Standard Quantum Limit with Four-Entangled Photons}.
\newblock {\em Science}, 316(5825):726--729, 2007.

\bibitem{ur-lasphys-15-146}
A.~B. U'Ren, C.~Silberhorn, R.~Erdmann, K.~Banaszek, W.~P. Grice, I.~A.
  Walmsley, and M.~G. Raymer.
\newblock Generation of pure-state single-photon wavepackets by conditional
  preparation based on spontaneous parametric downconversion.
\newblock {\em Laser Physics}, 15:146, 2005.

\bibitem{so-jmo-54-467}
A.~Soujaeff, S.~Takeuchi, K.~Sasaki, T.~Hasegawa, and M.~Matsui.
\newblock Heralded single photon source at 1550 nm from pulsed parametric down
  conversion.
\newblock {\em Journal of Modern Optics}, 54:467--471, 2007.

\bibitem{ga-oe-15-14870}
K.~{Garay-Palmett}, H.~J. {McGuinness}, O.~{Cohen}, J.~S. {Lundeen},
  R.~{Rangel-Rojo}, A.~B. {U'Ren}, M.~G. {Raymer}, C.~J. {McKinstrie},
  S.~{Radic}, and I.~A. {Walmsley}.
\newblock {Photon pair-state preparation with tailored spectral properties by
  spontaneous four-wave mixing in photonic-crystal fiber}.
\newblock {\em Optics Express}, 15:14870, 2007.

\bibitem{gr-pra-64-063815}
W.~P. {Grice}, A.~B. {U'ren}, and I.~A. {Walmsley}.
\newblock {Eliminating frequency and space-time correlations in multiphoton
  states}.
\newblock {\em Phys. Rev. A}, 64(6):063815, 2001.

\bibitem{agrawal}
G.P. Agrawal.
\newblock {\em Nonlinear fiber optics}.
\newblock Academic Press, 1995.

\bibitem{po-optex-11-22}
T.~Hedley J. Knight T. Birks P.~Russell J.~Pottage, David~Bird and P.~Roberts.
\newblock Robust photonic band gaps for hollow core guidance in pcf made from
  high index glass.
\newblock {\em Opt. Express}, 11:2854--2861, 2003.

\bibitem{al-njp-8-67}
O.~Alibart, J.~Fulconis, G.~K.~L. Wong, S.~G. Murdoch, W.~J. Wadsworth, and
  J.~G. Rarity.
\newblock Photon pair generation using four-wave mixing in a microstructured
  fibre: theory versus experiment.
\newblock {\em New J. Phys}, 8:67, 2006.

\bibitem{ho-prl-59-2044}
C.~K. Hong, Z.~Y. Ou, and L.~Mandel.
\newblock Measurement of subpicosecond time intervals between two photons by
  interference.
\newblock {\em Phys. Rev. Lett.}, 59:2044--2046, 1987.

\bibitem{co-prl-102-123603}
Offir Cohen, Jeff~S. Lundeen, Brian~J. Smith, Graciana Puentes, Peter~J.
  Mosley, and Ian~A. Walmsley.
\newblock Tailored photon-pair generation in optical fibers.
\newblock {\em Phys. Rev. Lett.}, 102(12):123603, 2009.

\bibitem{mo-njp-10-093011}
P.~J. {Mosley}, J.~S. {Lundeen}, B.~J. {Smith}, and I.~A. {Walmsley}.
\newblock {Conditional preparation of single photons using parametric
  downconversion: a recipe for purity}.
\newblock {\em New J.\ Phys.}, 10(9):093011, 2008.

\bibitem{hecht}
E.~Hecht and A.~Zajac.
\newblock {\em Optics}.
\newblock Addision-Wesley, Reading, MA, 1974.

\bibitem{br-prl-95-010501}
Daniel~E. Browne and Terry Rudolph.
\newblock Resource-efficient linear optical quantum computation.
\newblock {\em Phys. Rev. Lett.}, 95(1):010501, 2005.

\bibitem{br-prl-81-5932}
H.-J. Briegel, W.~D\"ur, J.~I. Cirac, and P.~Zoller.
\newblock Quantum repeaters: The role of imperfect local operations in quantum
  communication.
\newblock {\em Phys. Rev. Lett.}, 81:5932--5935, 1998.

\bibitem{he-nat-465-1052}
Morgan~P. Hedges, Jevon~J. Longdell, Yongmin Li, and Matthew~J. Sellars.
\newblock Efficient quantum memory for light.
\newblock {\em Nature}, 465:1052--1056, 2010.

\bibitem{li-prl-94-053601}
Xiaoying Li, Paul~L. Voss, Jay~E. Sharping, and Prem Kumar.
\newblock Optical-fiber source of polarization-entangled photons in the 1550 nm
  telecom band.
\newblock {\em Phys. Rev. Lett.}, 94(5):053601, 2005.

\bibitem{fu-prl-99-120501}
J.~Fulconis, O.~Alibart, J.~L. O'Brien, W.~J. Wadsworth, and J.~G. Rarity.
\newblock Nonclassical interference and entanglement generation using a
  photonic crystal fiber pair photon source.
\newblock {\em Physical Review Letters}, 99(12):120501, 2007.

\bibitem{fe-opx-15-23}
Alessandro Fedrizzi, Thomas Herbst, Andreas Poppe, Thomas Jennewein, and Anton
  Zeilinger.
\newblock A wavelength-tunable fiber-coupled source of narrowband entangled
  photons.
\newblock {\em Optics Express}, 15(23):15377--15386, 2007.

\bibitem{jeremiethesis}
Jeremie Fulconis.
\newblock {\em Fibre Source of Photon Pairs for Quantum Information}.
\newblock University of Bristol PhD Thesis, 2006.

\bibitem{be-phys-1964}
J.S. Bell.
\newblock On the einstein-podolsky-rosen paradox.
\newblock {\em Physics}, 1:195--200.

\bibitem{cl-prl-23-880}
J.F. Clauser, M.A. Horne, A.~Shimony, and R.A. Holt.
\newblock Proposed experiment to test local hidden variable theories.
\newblock {\em Phys. Rev. Lett.}, 23:880, 1969.

\bibitem{ja-pra-64-052312}
Daniel F.~V. James, Paul~G. Kwiat, William~J. Munro, and Andrew~G. White.
\newblock Measurement of qubits.
\newblock {\em Phys. Rev. A}, 64:052312, 2001.

\bibitem{ba-pra-61-010304}
K.~Banaszek, G.~M. D'Ariano, M.~G.~A. Paris, and M.~F. Sacchi.
\newblock Maximum-likelihood estimation of the density matrix.
\newblock {\em Phys. Rev. A}, 61:010304, 1999.

\bibitem{stephaneproject}
Stephane Altazin.
\newblock {\em Masters Thesis}.
\newblock University of Bristol, 2007.

\bibitem{ra-prl-65-1348}
J.~G. Rarity, P.~R. Tapster, E.~Jakeman, T.~Larchuk, R.~A. Campos, M.~C. Teich,
  and B.~E.~A. Saleh.
\newblock Two-photon interference in a mach-zehnder interferometer.
\newblock {\em Phys. Rev. Lett.}, 65(11):1348--1351, 1990.

\bibitem{bo-nt-390-575}
Dik Bouwmeester, Jian-Wei Pan, Klaus Mattle, Manfred Eibl, Harald Weinfurter,
  and Anton Zeilinger.
\newblock Experimental quantum teleportation.
\newblock {\em Nature}, 390:575--579, 1997.

\bibitem{pi-pra-64-062311}
T.~B. Pittman, B.~C. Jacobs, and J.~D. Franson.
\newblock Probabilistic quantum logic operations using polarizing beam
  splitters.
\newblock {\em Phys. Rev. A}, 64:062311, 2001.

\bibitem{cl-pra-79-030303}
A.~S. Clark, J.~Fulconis, J.~G. Rarity, W.~J. Wadsworth, and J.~L. O'Brien.
\newblock All-optical-fiber polarization-based quantum logic gate.
\newblock {\em Phys. Rev. A}, 79(3):030303, 2009.

\bibitem{pi-pra-68-032316}
T~B Pittman, M~J Fitch, B~C Jacobs, and J~D Franson.
\newblock Experimental controlled-not logic gate for single photons in the
  coincidence basis.
\newblock {\em Phys. Rev. A}, 68:032316, 2003.

\bibitem{ob-nat-426-264}
J.~L. O'Brien, G.~J. Pryde, A.~G. White, T.~C. Ralph, and D.~Branning.
\newblock Demonstration of an all-optical quantum controlled-\textsc{NOT} gate.
\newblock {\em Nature}, 426(6964):264--267, 2003.

\bibitem{ob-prl-93-080502}
J.~L. O'Brien, G.~J. Pryde, A.~Gilchrist, D.~F.~V. James, N.~K. Langford, T.~C.
  Ralph, and A.~G. White.
\newblock Quantum process tomography of a controlled-\textsc{NOT} gate.
\newblock {\em Phys. Rev. Lett.}, 93(8):080502, 2004.

\bibitem{ra-pra-65-062324}
T.~C. Ralph, N.~K. Langford, T.~B. Bell, and A.~G. White.
\newblock Linear optical controlled-{NOT} gate in the coincidence basis.
\newblock {\em Phys. Rev. A}, 65:062324, 2001.

\bibitem{la-prl-95-210504}
N.~K. Langford, T.~J. Weinhold, R.~Prevedel, K.~J. Resch, A.~Gilchrist, J.~L.
  O'Brien, G.~J. Pryde, and A.~G. White.
\newblock Demonstration of a simple entangling optical gate and its use in
  bell-state analysis.
\newblock {\em Phys. Rev. Lett.}, 95(21):210504, 2005.

\bibitem{ki-prl-95-210505}
Nikolai Kiesel, Christian Schmid, Ulrich Weber, Rupert Ursin, and Harald
  Weinfurter.
\newblock Linear optics controlled-phase gate made simple.
\newblock {\em Physical Review Letters}, 95(21):210505, 2005.

\bibitem{ra-prl-86-5188}
Robert Raussendorf and Hans~J. Briegel.
\newblock A one-way quantum computer.
\newblock {\em Phys. Rev. Lett.}, 86:5188--5191, 2001.

\end{thebibliography}
\bibliographystyle{unsrt}

\end{document}